\documentclass[preprint, review,3p,times, 10pt]{elsarticle}

\usepackage{amsmath,amssymb,amsfonts}
\usepackage{algorithmic}
\usepackage{graphicx}

\usepackage[utf8]{inputenc}
\usepackage{tabularx}   
\usepackage{float}      
\usepackage{booktabs}   
\usepackage{array}
\usepackage{xcolor}
\usepackage{soul}
\sethlcolor{green!20}

\usepackage{textcomp}
\usepackage{comment}
\usepackage{footnote}
\usepackage{threeparttable}
\usepackage{romannum}
\usepackage{longtable}
\usepackage{lscape}
\usepackage{soul}
\usepackage{array, makecell}
\usepackage{multirow}
\usepackage{multicol}
\usepackage{listings}
\usepackage{xcolor}

\usepackage{booktabs}
\usepackage{array}
\usepackage{amssymb} 

\usepackage{tikz}
\usepackage{pgfplots}
\usepackage{pgf-pie}
\usepackage{longtable}

\usepackage[hyphens]{url}
\usepackage{hyperref}
\hypersetup{colorlinks=true,breaklinks=true}

\journal{Elsevier}
\pgfplotsset{compat=1.18}
\begin{document}

\begin{frontmatter}
\title{LLaMA-XR: A Novel Framework for Radiology Report Generation using LLaMA and QLoRa Fine-Tuning}

\author[1]{Md. Zihad Bin Jahangir}
\ead{2018200000015@seu.edu.bd}

\author[2]{Muhammad Ashad Kabir \corref{correspondingauthor}}
\cortext[correspondingauthor]{Corresponding author: Charles Sturt University, Panorama Ave, Bathurst, NSW 2795.}
\ead{akabir@csu.edu.au}

\author[3]{Sumaiya Akter}
\ead{sumaiya.akter.cse@ulab.edu.bd}

\author[1]{Israt Jahan}
\ead{2020100000010@seu.edu.bd}

\author[4]{Minh Chau}
\ead{schau@csu.edu.au}

\affiliation[1]{organization={Department of Computer Science and Engineering, Southeast University}, city={Dhaka}, postcode= {1215}, country={Bangladesh}}

\affiliation[2]{organization={School of Computing, Mathematics and Engineering, Charles Sturt University}, city={Bathurst}, state={NSW}, postcode={2795}, country={Australia}}


\affiliation[3]{organization={Department of Computer Science and Engineering, University of Liberal Arts Bangladesh}, city={Dhaka}, postcode={1207}, country={Bangladesh}}

\affiliation[4]{organization={Medical Imaging Group, School of Dentistry and Medical Sciences, Charles Sturt University}, city={Wagga Wagga}, state={NSW}, postcode={2650}, country={Australia}}

\begin{abstract}
\textbf{Background:} The goal of automated radiology report generation is to help radiologists in their task of creating descriptive reports from chest radiographs. However, the process of creating coherent and contextually accurate reports has been challenging, mainly due to the intricacies of medical language and the need to correlate visual data with textual descriptions.\\
\textbf{Methods:} In this paper, we present \textit{LLaMA-XR}, a novel framework that integrates Meta LLaMA 3.1 Large Language Model with DenseNet-121-based image embeddings and Quantized Low-Rank Adaptation (QLoRA) fine-tuning. LLaMA-XR achieves improved coherence and clinical accuracy while maintaining computational efficiency. This efficiency is driven by an optimization strategy that enhances parameter utilization and reduces memory overhead, enabling faster report generation with lower computational resource demands.\\
\textbf{Results:}
The experiment conducted on the IU X-ray dataset demonstrates that LLaMA-XR outperforms a range of state-of-the-art methods. Our model achieves a ROUGE-L score of 0.433 and a METEOR score of 0.336, establishing new performance benchmarks in the domain.\\
\textbf{Conclusions:}
These results underscore LLaMA-XR's potential as an effective artificial intelligence system for automated radiology reporting, offering enhanced clinical utility and reliability.
\end{abstract}

\end{frontmatter}

\section{Introduction}

Medical imaging plays a crucial role in diagnosing diseases by providing detailed visual representations of internal structures~\citep{wang2018tienet}. Among various radiological imaging modalities, chest X-ray (CXR) is one of the most commonly used techniques for diagnosing a wide range of pulmonary and cardiovascular conditions~\citep{mr2017acquired}. Given its widespread use in clinical practice, CXR reports are essential for accurately interpreting specific abnormalities and guiding treatment decisions~\citep{bahl2020interpretation}. However, generating these reports is a labor-intensive process that demands meticulous analysis of radiological images, making it both time-consuming and susceptible to errors~\citep{liu2019clinically}.

With the advancement of deep neural networks, automatic radiology report generation has emerged as a promising tool to assist radiologists~\citep{sloan2024automated}. This approach requires models to generate accurate, coherent, and medically precise reports based on radiological images, significantly reducing workload and improving efficiency. Early studies~\citep{you2016image, liu2018simnet} in automated radiology report generation were largely inspired by recurrent neural network (RNN)-based image captioning architectures. While these methods achieved initial success, they struggled with handling long and complex sentences~\citep{vinyals2015show, iftikhar2022iqra}, which are crucial for generating detailed and clinically relevant reports. 

To overcome these challenges, recent advances have focused on the use of large language models (LLMs), such as GPT~\citep{radford2018improving}and LLaMA~\citep{touvron2023llama}, which have benefited from extensive pre-training on diverse datasets and demonstrate superior performance in generating radiology reports~\citep{ranjit2023retrieval, adams2023leveraging, buckley2311accuracy, liu2023deid, kung2023performance, tanida2023interactive, MedicalGPT, li2025s}. In chest X-ray radiology, the latest multimodal foundation models aim to combine visual understanding with clinical language generation, covering abnormal assessment and report generation~\citep{chen2024visionlanguagefoundationmodelenhance, lee2025cxrllava}. Those LLM-based approaches have shown significant promise in capturing intricate medical contexts and producing high-quality reports. However, earlier versions of these models, such as GPT-2 (up to 1.5B parameters) and BERT-Large (340M parameters), often lacked the capacity to effectively handle the complexity of medical data due to their relatively smaller scale. This limitation underscores the need for a more advanced framework to better align with the demands of medical report generation.

In this study, we propose a novel framework for automated chest X-ray report generation that addresses the challenge of effectively handling the complexity of medical data. Our approach takes advantage of the LLaMA 3.1 8B parameter model~\citep{dubey2024llama}, which offers superior performance in generating coherent, contextually rich, and medically precise reports. Compared to its predecessors, LLaMA 3.1 benefits from enhanced training data and a significantly larger parameter scale, enabling it to better capture the nuances of medical language~\citep{nicolson2023improving, tao2024memory}. To extract visual features from CXR images, several techniques can be employed, including convolutional neural networks (CNNs) such as DenseNet, ResNet, and Vision Transformers. In this study, we utilize DenseNet-121~\citep{cohen2022torchxrayvision}, a CNN architecture known for its ability to reuse features across layers and provide detailed image embeddings~\citep{nguyen2021automated, nicolson2023improving}. However, this choice is flexible, and other architectures, such as ResNet, EfficientNet, or Vision Transformers, can be incorporated depending on specific requirements. By combining DenseNet-121 with LLaMA 3.1, our framework effectively bridges the gap between visual data and textual report generation.

Furthermore, we employ fine-tuning techniques such as QLoRA~\citep{dettmers2024qlora}, which enables efficient adaptation of large language models by significantly reducing memory requirements through low-rank adaptation, allowing us to fine-tune LLaMA 3.1 on medical datasets without the high computational cost of full fine-tuning.
We use Supervised Fine-Tuning (SFT) to adapt LLaMA 3.1 to medical datasets, ensuring that the generated reports are accurate and clinically relevant. For this study, we used the IU X-ray dataset~\citep{demner2016preparing}, a widely used benchmark dataset that contains chest X-ray images paired with the corresponding radiology reports. This dataset enables effective model training and evaluation by providing diverse medical cases with detailed annotations. This fine-tuning process not only improves the quality of report generation but also reduces the time and computational resources required for training. Our framework is designed to be scalable and efficient, making it suitable for real-world applications where accuracy and interpretability are paramount.
The key contributions of this study are as follows:

\begin{itemize}
\item We propose a novel framework for automated chest X-ray report generation that leverages LLaMA 3 as the core language model, enhanced through parameter-efficient QLoRA and supervised fine-tuning (SFT).

\item We perform an extensive comparison with state-of-the-art (SOTA) methods, demonstrating superior performance in ROUGE-L and METEOR metrics.

\item We further assess the clinical quality of the generated reports, demonstrating strong factual accuracy, semantic alignment, and clinical relevance.

\end{itemize}

\section{Related work}

Automated radiology report generation has garnered significant attention in recent years, with a wide array of deep learning architectures being proposed to enhance diagnostic accuracy, linguistic coherence, and computational efficiency. This section reviews the evolution of this field, focusing on methods applicable to generate chest X-ray report. We categorize recent work into four main classes: graph-based models, reward-based learning approaches, transformer-based architectures, and large language models (LLMs). 

\subsection{CNN-RNN Based Methods}
The roots of radiology report generation lie in image captioning, a foundational computer vision task designed to produce descriptive sentences for visual input. Early models typically utilized CNN-RNN frameworks, where Convolutional Neural Networks (CNNs) extracted spatial features and Recurrent Neural Networks (RNNs) generated the corresponding captions~\citep{islam2021exploring, suresh2022image, zhang2024review}. However, these approaches often struggled with long-range dependencies and lacked contextual depth. One of the earliest tailored approaches for radiology reports is Tm-HRNN~\citep{yin2019automatic}, which introduced a topic modeling-based hierarchical RNN framework. It used DenseNet for image feature extraction, followed by a CNN-RNN-based decoder that generates reports in a sentence-by-sentence fashion, guided by topic vectors. This hierarchical structure mimicked how radiologists organize findings into coherent paragraphs, laying an early foundation for structure-aware generation. Despite its innovative structure, the model relied on heuristic topic assignments and lacked explicit medical knowledge modeling.

To address these limitations, later enhancements like anchor-caption mechanisms~\citep{xu2021towards}, human-like control techniques~\citep{chen2021human}, and self-attention networks~\citep{tran2020transform} were introduced. While effective in general contexts, their reliance on object annotations made them less suitable for medical imaging tasks, where such annotations are scarce or unavailable. This led to the development of medical imaging captioning models that employed image embeddings combined with large language models (LLMs) to extract richer semantic features without extensive supervision. The transition from captioning to report generation marked a shift from surface-level description to clinical reasoning and structured reporting, requiring models to understand complex visual cues and their clinical implications.


\subsection{Graph-Based and Reward-Based Methods}
To better capture the interdependencies between anatomical entities and clinical findings, researchers have turned to graph neural networks (GNNs). These models encode relationships between observed pathologies and anatomical locations to improve report coherence. For example,~\citet{zhang2020radiology} introduced SentSAT+KG, which combines DenseNet-121 features with a knowledge graph to guide the generation process. Similarly, \citet{li2023dynamic} proposed DCL, a ViT-based approach that leverages dynamic graphs to represent complex entity relationships for improved interpretability and performance. However, these models require the construction and maintenance of complex medical knowledge graphs, which can be labor-intensive and often depend on domain expertise. Moreover, their performance is highly sensitive to the quality and completeness of the graph, making them less robust in diverse clinical scenarios.

To optimize report generation beyond traditional supervised objectives, reinforcement learning (RL) has been explored. These methods employ task-specific reward functions that promote diagnostic relevance and linguistic quality. Notably,~\citet{li2018hybrid} proposed the H-Agent model, leveraging DenseNet-based image features with a reward-based training strategy. Building on this,~\citet{jing2020show} introduced CMAS-RL, which uses ResNet-50 as the backbone and incorporates domain-specific rewards to refine output quality. Further enhancing this line of research,~\citet{qin2022reinforced} presented CMM+RL, which applies a more powerful ResNet-101 encoder and a reinforcement learning framework to improve factual correctness and report coherence. Nonetheless, reinforcement learning-based models often face challenges such as unstable training dynamics, sensitivity to reward function design, and high computational overhead, which hinder their practical deployment in real-world clinical systems.

\subsection{Transformer and Large Language Models Based Methods}
The transformative success of self-attention mechanisms in NLP has inspired numerous transformer-based models for radiology. These models can better contextualize spatial and semantic features across the image-report pipeline. For example, the original Transformer model~\citep{vaswani2017attention} laid the foundation for this architecture class. R2Gen~\citep{chen2020generating}, PPKED~\citep{liu2021exploring}, and R2GenCMN~\citep{chen2022cross} adopted custom transformer designs on top of ResNet-based image encoders to generate structured reports. Models such as SGF~\citep{li2022self}, PGT~\citep{yan2022prior}, JPG~\citep{you2022jpg}, and ITA~\citep{wang2022inclusive} further refine this paradigm by integrating domain-specific priors, knowledge embeddings, or training schemes such as curriculum learning, as used in CMCL~\citep{liu2022competence}. Variational approaches like VTI~\citep{najdenkoska2021variational} and vision-language alignment methods such as UAR~\citep{li2023unify} demonstrate the versatility of transformer-based architectures in balancing generation fidelity and clinical accuracy. However, these models often struggle with generating detailed and accurate descriptions, especially when handling complex sentence structures and mitigating data biases, largely because they are not pre-trained on large-scale, diverse datasets to effectively capture the nuanced relationships between visual features and textual descriptions.

The application of LLMs, such as GPT and LLaMA, has revolutionized medical report generation by providing powerful pre-trained language priors. These models can generate coherent, factually accurate reports using multimodal cues.~\citet{nicolson2023improving} introduced CvT-DistilGPT2, combining CvT-21 image features with a lightweight DistilGPT2 language model to enhance generation efficiency while preserving quality. This model exemplifies how compact transformer variants can be used effectively in resource-constrained environments. Another notable approach is MCSAM by~\citet{tao2024memory}, which uses MedCLIP for visual encoding and integrates it with BERT for report generation. This memory-compressed alignment model leverages semantic consistency across image-text pairs, highlighting the value of aligning cross-modal embeddings. However, despite its efficiency, the reduced model size limits its capacity to capture complex medical terminologies and relationships, potentially affecting the depth and accuracy of the generated reports.


In summary, existing research on the generation of chest radiographs has mainly relied on earlier transformer models, resulting in notable performance improvements. However, these models still encounter limitations in generating highly relevant and medically precise reports. A key gap in the literature is the under utilization of larger language models, such as LLaMA-3, which have the potential to enhance both the accuracy and contextual understanding necessary for clinical applications. We address this gap by leveraging the LLaMA-3.1 8B model, which offers improved performance and better contextual comprehension to generate more accurate and medically relevant reports.

\section{Material and Methods}

\subsection{LLaMA-XR Framework}

\begin{figure}[h!]
    \centering
    \includegraphics[width=1\textwidth]{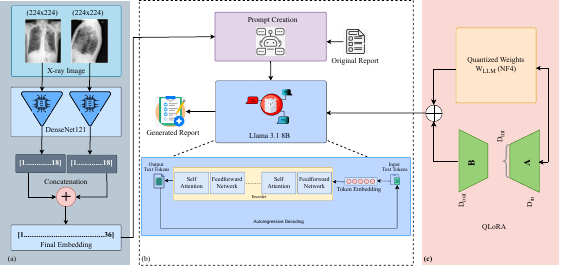}
    \caption{Overview of the proposed framework for automated radiology report generation. Chest X-ray images from anteroposterior (AP) and lateral (LAT) views are processed using DenseNet-121 to extract visual embeddings, which are concatenated and combined with corresponding radiology reports using Alpaca-style prompt formatting. The prompts are then provided to Llama 3.1 8B to generate the final radiology report.}
    \label{fig:model_architecture}
\end{figure}

Our framework involves two primary components: one for visual information extraction from CXRs and the other for generating reports using LLMs. For visual information extraction, we utilized DenseNet-121, and for generating reports, we employed LLaMA 3. Figure~\ref{fig:model_architecture} provides an illustration of this framework.

The ``DenseNet121-res224-all" model~\citep{cohen2022torchxrayvision} was pre-trained on a broad collection of public CXR datasets that includes the IU X-ray dataset. Although the visual extractor is kept completely frozen (no gradients are back-propagated) and is used only to produce fixed 36-dimensional embeddings, we acknowledge that pre-training on IU X-ray may introduce a potential source of bias. In particular, the model may generate higher-quality embeddings for images drawn from a familiar distribution, which could slightly inflate performance compared to entirely unseen datasets. This limitation is consistent with prior work using large-scale pre-trained CXR models and should be considered when interpreting results. In the model name, DenseNet-121 refers to the architecture, res224 indicates $224 \times 224$ pixels resolution, and `all' signifies that the model was trained on all available chest X-ray datasets. DenseNet-121 is a convolutional neural network known for its dense connectivity pattern, which enables feature reuse across layers and promotes the learning of compact and informative image representations~\citep{nguyen2021automated, nicolson2023improving}. This model extracts detailed visual features from chest radiographs, outputs confidence scores for 18 different medical conditions: Atelectasis, Cardiomegaly, Consolidation, Edema, Effusion, Emphysema, Enlarged Cardiomediastinum, Fibrosis, Fracture, Hernia, Infiltration, Lung Lesion, Lung Opacity, Mass, Nodule, Pleural Thickening, Pneumonia, and Pneumothorax. These scores serve as semantically rich embeddings for the report generation module, guiding the language model in producing clinically meaningful narratives. We chose to use a small 36-dimensional vector of 18-condition confidence scores instead of larger high-dimensional vectors. This allows us to have clinically interpretable priors, reduce the dimensionality of the input for the LLM, and align with radiology terminology. 

\begin{figure}[!ht]
    \centering
    \includegraphics[width=1\textwidth]{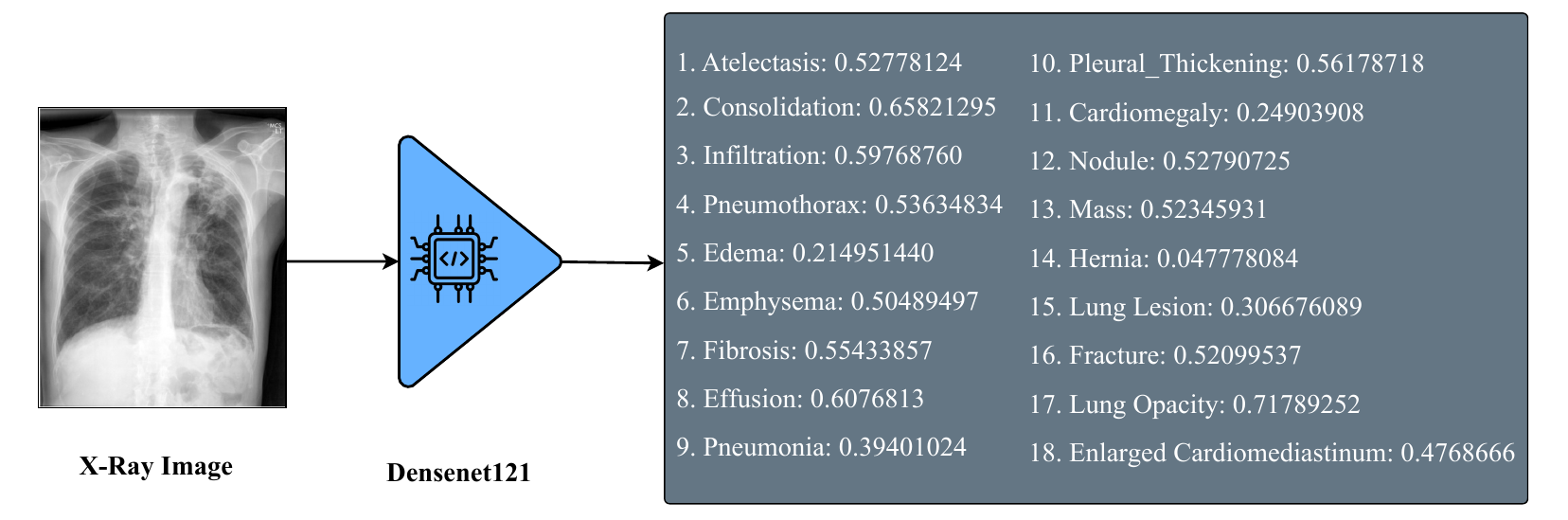}
    \caption{Illustration of the ``DenseNet121-res224-all" output classes.}
    \label{fig:densnetOutput}
\end{figure}

\begin{figure}[!ht]
    \centering
    \includegraphics[width=1\textwidth]{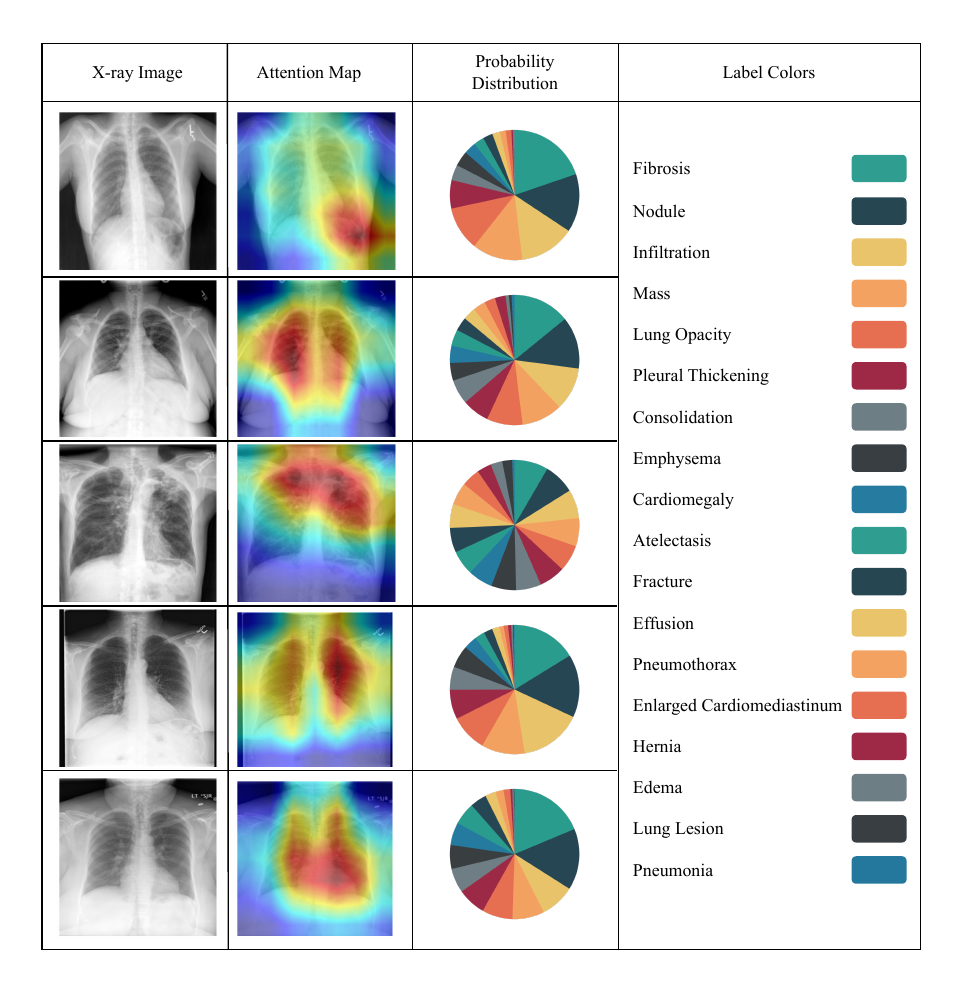}
    \caption{The heatmap highlights the critical regions in a chest X-ray image that influenced the Densenet-121 model's classification decision. Warmer colors indicate higher importance in the model's feature extraction process. The probability distribution illustrates the confidence score of each abnormality, and 18 colors represent 18 abnormalities.}
    \label{fig:heatmap}
\end{figure}

This approach has the benefits of reducing computational costs without compromising on strong semantic performance. We also have plans for an ablation study where we compare against late-layer CNN/ViT embeddings, and the modularity makes this very accessible. Although DenseNet-121 was selected due to its strong performance and efficient parameter usage in medical imaging tasks, our framework is modular and flexible, allowing alternative architectures such as ResNet, EfficientNet, or Vision Transformers to be substituted based on specific needs or experimental goals.
Figure~\ref{fig:densnetOutput} illustrates the output classes of the multi-label classification, as determined by the pre-trained DenseNet classifier. Figure~\ref{fig:heatmap} then adds a heatmap to the chest X-ray image, indicating the parts of the image that were considered most relevant during the classification process. These heatmaps are based on the DenseNet feature extractor and should not be considered as an explanation of the report generation process of the LLaMA component. These visualizations enhance interpretability by linking model decisions to specific image regions, thereby supporting transparent and informed clinical use.

LLaMA 3's architecture is a part of the LLaMA series, which are transformer-based models. The LLaMA series models are trained on vast amounts of textual data to predict the next token in a sequence, making them highly effective for tasks such as text generation, translation, and summarization. Specifically, the ``LLaMA 3.1 8B"~\citep{dubey2024llama} model has 8 billion parameters, which makes it capable of understanding and generating complex and contextually accurate text. The architecture of LLaMA 3 is illustrated in Figure~\ref{fig:model_architecture}, consisting of multiple layers of transformer blocks, each containing \textit{self-attention} mechanisms and \textit{ feedforward} neural networks. These layers enable the model to capture long-range dependencies in the text, making it adept at understanding context and generating coherent and relevant text based on the given input. The model uses a tokenization process to convert input text into tokens, which are then processed through the transformer layers. The output is a sequence of tokens converted back into readable text, resulting in detailed and contextually accurate reports.

For generating reports, we used the ``LLaMA 3.1 8B 4-bit quantized" model~\citep{dubey2024llama}, implemented via the Unsloth\footnote{\url{https://unsloth.ai/}} framework alongside two fine-tuning strategies: SFT and QLoRA~\citep{dettmers2024qlora}. The LLaMA 3.1 8B model was selected for its strong performance in language understanding and generation, offering a robust balance between model size and capability. The 8B parameter scale is large enough to capture complex medical semantics while remaining computationally feasible. The 4-bit quantization significantly reduces memory use and accelerates inference, but does so with little impact to model performance, and so it is a good candidate for practical use, especially in resource-constrained settings. Unsloth was used because of its highly optimized implementation for quantized models, enabling faster and more memory-efficient training. SFT improves the ability of the model to generalize in domain-specific tasks by training it on labeled radiology report data. Meanwhile, QLoRA enables low-rank adaptation of a quantized base model, updating only a small, learnable subset of parameters. This approach improves efficiency, reduces overfitting, and reduces overall training cost. Together, this configuration allows us to fine-tune the model up to two times faster and with approximately 60\% less memory usage~\citep{dettmers2024qlora}, making the entire system scalable and accessible for wider real-world applications in the generation of medical reports.

\subsection{QLoRA}

The QLoRA framework~\citep{dettmers2024qlora} is a significant advancement over the Low-Rank Adaptation (LoRA)~\citep{hu2021lora} technique, specifically designed to make fine-tuning LLMs more efficient in terms of memory usage and computational cost. QLoRA builds upon the core concept of LoRA, which freezes the pre-trained weight matrix, $W_0 \in \mathbb{R}^{d \times k}$, during fine-tuning to prevent computationally expensive updates. Instead, LoRA introduces two trainable low-rank matrices, $A \in \mathbb{R}^{d \times r}$ and $B \in \mathbb{R}^{r \times k}$, where $r$ is the rank hyperparameter, representing a trade-off between efficiency and model expressiveness. The weight updates in QLoRA are similarly represented as in Equation~\ref{eq:delta_W}, offering a low-rank approximation of the necessary adjustments to the original weights. 

\begin{equation}
\Delta W = A B,
\label{eq:delta_W}
\end{equation}
where $A \in \mathbb{R}^{d \times r}$ and $B \in \mathbb{R}^{r \times k}$ represent the low-rank matrices, and $\Delta W$ is the update to the original weights.

At inference, the combined weight matrix $W$ is computed by merging the frozen pre-trained weights and the learned updates, as shown in Equation~\ref{eq:merged_weights}.
\begin{equation}
W = W_0 + \Delta W.
\label{eq:merged_weights}
\end{equation}
where $W_0$ is the frozen pre-trained weight matrix, and $\Delta W$ is the learned update from the low-rank matrices.

The forward pass in QLoRA is expressed as:
\begin{equation}
h = W_0 x + A B x,
\label{eq:forward_pass}
\end{equation}
where $x$ is the input data, and $h$ is the hidden representation. This decomposition enables the model to retain the general knowledge encoded in $W_0$ while incorporating task-specific information through $A$ and $B$.

QLoRA~\citep{dettmers2024qlora} further improves upon LoRA~\citep{hu2021lora} by introducing 4-bit quantization for the pre-trained weight matrix $W_0$. Quantization is a technique that reduces the precision of numerical representations, enabling computations with lower memory and faster execution speeds. In QLoRA, the pre-trained weights are quantized to 4 bits, significantly reducing the memory footprint of the model by approximately 60\%. Despite the reduction in precision, this quantization minimally affects model accuracy because $W_0$ remains frozen and is complemented by the precise updates provided by the full-precision low-rank matrices $A$ and $B$. This hybrid approach ensures that task-specific adaptations are handled with high fidelity while general pre-trained knowledge is retained in a compressed form.

Another critical component is the rank hyperparameter ($r$) in the low-rank matrices. By adjusting $r$, practitioners can balance the trade-off between model performance and computational efficiency. A lower rank results in fewer parameters to update, which reduces memory and accelerates training, while a higher rank increases the model's capacity to learn complex task-specific patterns.

These innovations in QLoRA lead to substantial performance gains. By restricting updates to the smaller matrices $A$ and $B$ and leveraging quantized pre-trained weights, QLoRA achieves up to twice the training speed of traditional fine-tuning methods. The memory requirements are reduced by more than 60\%, enabling the fine-tuning of large-scale models such as LLaMA 3.1 8B~\citep{dubey2024llama} on hardware as limited as a single A100 GPU. This efficiency does not compromise the model's ability to produce high-quality, task-specific outputs, making QLoRA a practical solution for deploying LLMs in resource-constrained environments.

\subsection{Dataset Description}
We evaluated our approach using the IU-Xray~\citep{demner2016preparing} dataset, which consists of 7,470 chest X-ray images and 3,955 corresponding radiology reports. Each report is paired with both a frontal and a lateral X-ray image or contains one image only. In the IU X-ray dataset, approximately 12.4\% of cases contain only a single view. These single-view cases were proportionally distributed across the patient-level 70\% training, 10\% validation, and 20\% test splits, ensuring a consistent data distribution across all subsets. For such cases, the 18 confidence scores from the available view are duplicated to maintain the required 36-dimensional input representation. This approach preserves architectural consistency without introducing additional preprocessing complexity. Given the relatively small proportion of single-view cases, their overall impact on model performance is expected to be limited, and no significant deviation in training dynamics or evaluation metrics was observed. Table~\ref{fig:example_dataset} presents two patients' X-ray images from the dataset along with their associated reports.

\begin{table}[!htbp]
\centering
\caption{Examples from the IU-Xray dataset, showing chest X-ray images (frontal and lateral views) paired with their corresponding radiology reports.}
\label{fig:example_dataset}
\begin{tabular}{lp{10cm}}
\hline
 Radiology Image & Radiology Report \\ 
 \hline
\hline

\raisebox{-1\height}{\includegraphics[width=0.25\textwidth]{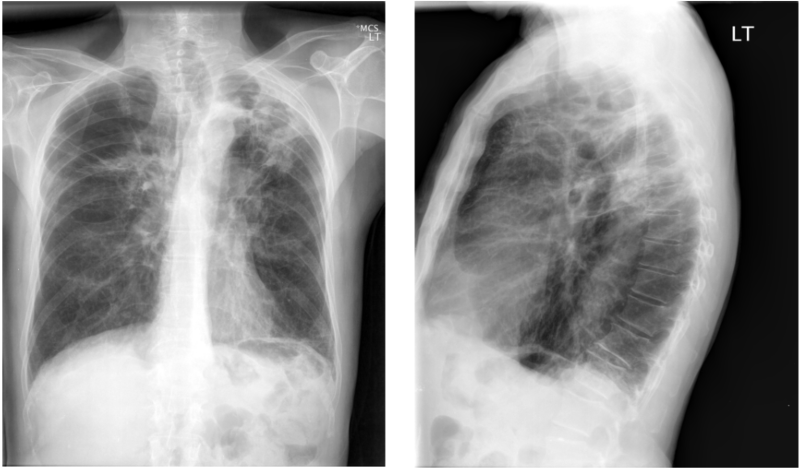}} & 
\textit{There are diffuse bilateral interstitial and alveolar opacities consistent with chronic obstructive lung disease and bullous emphysema. There are irregular opacities in the left lung apex that could represent a cavitary lesion in the left lung apex. There are streaky opacities in the right upper lobe, XXXX scarring. The cardiomediastinal silhouette is normal in size and contour. There is no pneumothorax or large pleural effusion}. \\ \hline

\raisebox{-1\height}{\includegraphics[width=0.25\textwidth]{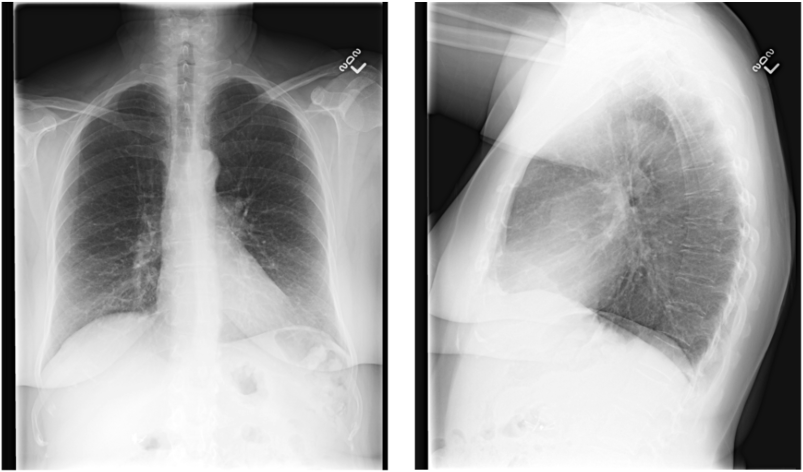}} & 
\textit{The cardiomediastinal silhouette and pulmonary vasculature are within normal limits. There is no pneumothorax or pleural effusion. There are no focal areas of consolidation. Cholecystectomy clips are present. Small T-spine osteophytes. There is biapical pleural thickening, unchanged from the prior. Mildly hyperexpanded lungs}. \\ \hline

\end{tabular}
\end{table}

\subsection{Dataset Preparation}
Our dataset preparation process is divided into two key steps. The first step focuses on preparing the data for the visual information extractor, which extracts useful features from the X-ray images. The second step involves using these extracted features to fine-tune the LLM for medical report generation.

 \paragraph{Image Preparation}
In the first step, we used DenseNet121-res224-all~\citep{cohen2022torchxrayvision} for the visual information extractor. To prepare the images for this model, we first resized all X-ray images to a consistent dimension of $224\times224$ pixels. This resizing ensures that the input images meet the required size for the DenseNet-121 model. After resizing, we normalized the images to scale the pixel values, making them consistent and aiding the model’s performance. Since X-ray images are typically grayscale, we also converted the images to a single color channel to ensure that only the relevant intensity information is used during feature extraction. Once processed, the images were fed into the DenseNet-121 network, which outputs a 36-dimensional vector representing classification scores for 18 different medical conditions across two views: frontal and lateral, as illustrated in Figure~\ref{fig:model_architecture}. These vectors are critical for the subsequent steps in the report generation process.

\begin{figure}[h!]
    \centering
    \includegraphics[width=1\textwidth]{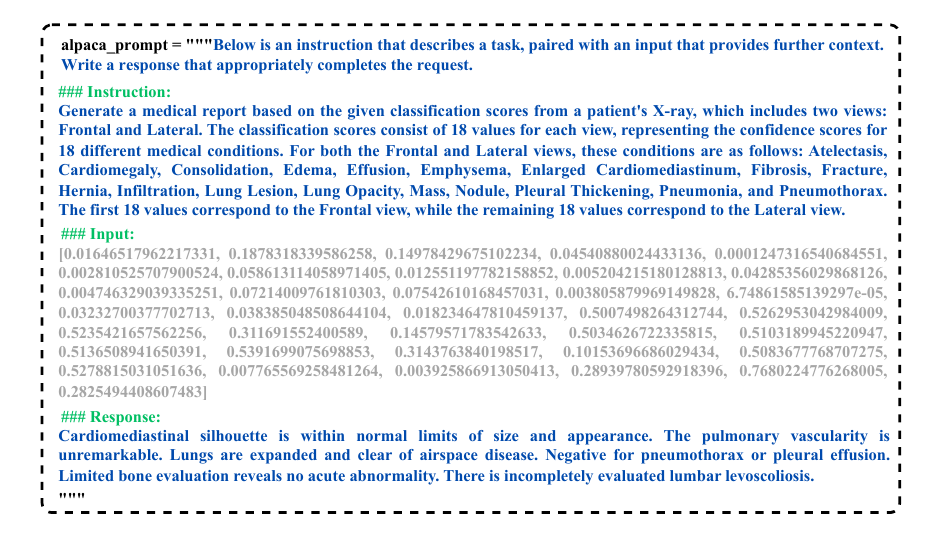}
    \caption{Example prompt for fine-tuning}
    \label{fig:example_prompt}
\end{figure}

 \paragraph{Prompt Generation}
 In the second step, we used the extracted visual information to construct a dataset for fine-tuning the LLM. For this purpose, we employed Unsloth\footnote{\url{https://unsloth.ai/}} and QLoRA~\citep{dettmers2024qlora} techniques, along with a structured Alpaca prompt system to format the input data appropriately. The Alpaca prompt format, as shown in Figure~\ref{fig:example_prompt}, begins with a general instruction that describes the task at hand. This is followed by a specific instruction that provides detailed context about the input data, which in this case consists of a 36-dimensional vector, 18 values from the frontal view and 18 values from the lateral view. The input part of the prompt contains these classification scores, while the corresponding response contains the relevant findings based on the medical conditions associated with the scores.

During the testing phase, the response portion of the prompt remains empty. This allows the model to generate the medical report based on the input data, as it will be expected to fill in the response with the appropriate findings for the given classification scores.

\subsection{Fine-Tuning LLaMA}
For fine-tuning the LLaMA 3.1 model, we used Google Colab equipped with an A100 GPU, which provided the computational resources necessary to efficiently handle large-scale fine-tuning tasks. The training process was implemented using the \texttt{SFTTrainer} from the \emph{trl} library. To achieve an effective balance between computational efficiency and gradient stability, the batch size was set to 8, with gradient accumulation steps of 4, effectively simulating a batch size of 32. The model was trained for three complete epochs, which was sufficient for convergence, given the size of the training dataset and the complexity of the task. A learning rate of $2 \times 10^{-6}$ was used, optimized using the AdamW 8-bit optimizer (\texttt{optim="adamw\_8bit"}), which significantly reduces memory overhead by quantizing the optimizer states without compromising performance. Mixed precision training (\texttt{fp16}) was enabled to take advantage of the A100 GPU's capabilities, accelerating training while reducing memory usage. The fine-tuning involved a total of 41,943,040 trainable parameters, enabling the model to adapt to the target domain while preserving its pre-trained knowledge.

A linear learning rate scheduler (\texttt{lr\_scheduler\_type="linear"}) was employed to ensure smooth and controlled adjustments to the learning rate during training, while warmup steps (\texttt{warmup\_steps=30}) facilitated a gradual ramp-up to the optimal learning rate, mitigating potential instabilities during the initial training phase. A weight decay of 0.01 was applied to regularize the model, preventing overfitting by penalizing large weight magnitudes. To further enhance training robustness, early stopping was implemented, thereby avoiding unnecessary computations and overfitting. Model checkpoints were periodically saved, with the best-performing model automatically restored at the end of training (\texttt{load\_best\_model\_at\_end=True}). Evaluations were conducted at regular intervals to track performance metrics and ensure the model was progressing as expected.

\subsection{Evaluation Metrics}
\paragraph{BLEU}
BiLingual Evaluation Understudy (BLEU)~\citep{papineni2002bleu} is a metric that compares human-generated output with machine-generated output based on n-gram. It evaluates how accurately the machine has generated the report by comparing it with the reference report. The BLEU score is defined in Equation~\ref{eq:bleu}, where \(BP\) is the brevity penalty as shown in Equation~\ref{eq:brevity_penalty}, \(p_n\) is the precision of \(n\)-grams given in Equation~\ref{eq:precision}, and \(w_n\) is the weights for \(n\)-grams as shown in Equation~\ref{eq:weights}.

\begin{equation}
\text{BLEU} = BP \cdot \exp \left( \sum_{n=1}^N w_n \log p_n \right)
\label{eq:bleu}
\end{equation}

\begin{equation}
BP = 
\begin{cases} 
1 & \text{if } c > r, \\
e^{(1 - \frac{r}{c})} & \text{if } c \leq r
\end{cases}
\label{eq:brevity_penalty}
\end{equation}

\begin{equation}
p_n = \frac{\text{Number of matching \(n\)-grams}}{\text{Total number of \(n\)-grams in candidate}}
\label{eq:precision}
\end{equation}

\begin{equation}
w_n = \frac{1}{N}
\label{eq:weights}
\end{equation}

\paragraph{ROUGE-L}
The ROUGE-L (Recall-Oriented Understudy for Gisting Evaluation - Longest Common Subsequence)~\citep{lin2004rouge} score is calculated based on the longest common subsequence (LCS) between a reference text and a candidate text. The ROUGE-L score considers precision, recall, and an F-measure based on the LCS and is defined in Equation~\ref{eq:rouge_l}. The precision and recall are computed using Equation~\ref{eq:precision_lcs} and Equation~\ref{eq:recall_lcs}, respectively. Here, \(\text{LCS}\) is the length of the longest common subsequence between the candidate and reference sequences, \(\beta\) is a weighting factor (commonly set to 1 to equally weight precision and recall), and \(F_1\) is the harmonic mean of precision and recall.

\begin{equation}
\text{ROUGE-L} = F_1 = \frac{(1 + \beta^2) \cdot \text{Precision} \cdot \text{Recall}}{\beta^2 \cdot \text{Precision} + \text{Recall}}
\label{eq:rouge_l}
\end{equation}

\begin{equation}
\text{Precision} = \frac{\text{LCS}}{\text{Length of the candidate sequence}}
\label{eq:precision_lcs}
\end{equation}

\begin{equation}
\text{Recall} = \frac{\text{LCS}}{\text{Length of the reference sequence}}
\label{eq:recall_lcs}
\end{equation}


\paragraph{METEOR}

The Metric for the Evaluation of Translation with Explicit ORdering (METEOR)~\citep{denkowski2011meteor} score evaluates the similarity between a candidate sentence and a reference sentence using, and \(F_\text{mean}\) (a harmonic mean of precision and recall) and penalty $P$,  as defined in Equation~\ref{eq:meteor}. Precision and recall are calculated using Equation~\ref{eq:meteor_precision} and Equation~\ref{eq:meteor_recall}, respectively. $F_\text{mean}$ is the harmonic mean of precision and recall controlled by the parameter $\alpha$, as shown in Equation~\ref{eq:meteor_fmean}. Furthermore, the penalty function to handle excessive fragmentation is defined in Equation~\ref{eq:meteor_penalty}.

\begin{equation}
\text{METEOR} = F_\text{mean} \cdot (1 - P)
\label{eq:meteor}
\end{equation}

\begin{equation}
\text{Precision} = \frac{\text{Number of matched unigrams}}{\text{Total unigrams in the candidate sentence}}
\label{eq:meteor_precision}
\end{equation}

\begin{equation}
\text{Recall} = \frac{\text{Number of matched unigrams}}{\text{Total unigrams in the reference sentence}}
\label{eq:meteor_recall}
\end{equation}

\begin{equation}
F_\text{mean} = \frac{\text{Precision} \cdot \text{Recall}}{\alpha \cdot \text{Precision} + (1 - \alpha) \cdot \text{Recall}}
\label{eq:meteor_fmean}
\end{equation}

\begin{equation}
P = \gamma \cdot \left( \frac{\text{Chunks}}{\text{Matched unigrams}} \right)^{\beta}
\label{eq:meteor_penalty}
\end{equation}
To fine-tune the METEOR metric, $\alpha$ is used, enhancing the correlation between human and machine translation evaluation. $\gamma$ and $\beta$ are parameters that control the severity of the penalty (typically $\gamma=0.5$ and $\beta=3$), $Chunks$ refers to contiguous matched segments in the candidate sentence, and \textit{Matched unigrams} is the number of unigram matches between the candidate and reference sentences.

The METEOR score penalizes fragmented matches to improve alignment quality and ensures a balanced combination of precision and recall.

\section{Results}
Table~\ref{tab:comparison} presents a comparative evaluation of our proposed method, LLaMA-XR, against state-of-the-art (SOTA) methods previously benchmarked on the IU X-ray dataset. LLaMA-XR demonstrates superior performance on ROUGE-L~\citep{lin2004rouge} and METEOR~\citep{denkowski2011meteor}, two widely used metrics that emphasize \textit{semantic} fidelity and \textit{syntactic} fluency. In particular, LLaMA-XR achieves a 4.34\% improvement in ROUGE-L over the state-of-the-art best result from the SGF method~\citep{li2022self}. This significant gain is primarily attributed to the incorporation of a larger pre-trained language model (LLM) within our architecture, which enhances the model’s ability to generate radiology reports that are both contextually relevant and clinically coherent. Moreover, LLaMA-XR surpasses the VTI \citep{najdenkoska2021variational} and UAR \citep{li2023unify} methods in METEOR metric by an impressive 54.13\%, underscoring its strength in capturing synonym variability and maintaining semantic alignment with the reference reports. Although LLaMA-XR performs slightly below some recent methods in terms of BLEU-4 scores, it is important to note that the BLEU metric~\citep{papineni2002bleu} primarily quantifies n-gram overlap, which, while effective for general language generation tasks, is less suited to evaluating the nuanced requirements of clinical text generation. In the context of radiology reporting, where semantic accuracy and clinical relevance are paramount, metrics like ROUGE-L and METEOR provide a more meaningful assessment of model quality.

\begin{table}[!htp] 
\centering 
\caption{Comparison with state-of-the-art methods. All results are quoted from the respective published studies, and boldface indicates the highest reported score for each metric.} \label{tab:comparison} 
\renewcommand{\arraystretch}{0.9}
\begin{tabular}{lcccc} 
\hline Method & Year & BLEU-4 & ROUGE-L & METEOR \\
\hline 
\hline
Transformer \citep{vaswani2017attention} & 2017 & 0.137 & 0.335 & 0.172 \\
H-Agent \citep{li2018hybrid} & 2018 & 0.151 & 0.322 & - \\ 
CMAS-RL \citep{jing2020show} & 2019 & 0.154 & 0.362 & - \\
Tm-HRNN \citep{yin2019automatic} & 2019 & 0.154 & 0.344 & 0.175 \\
SentSAT+KG \citep{zhang2020radiology} & 2020 & 0.147 & 0.367 & - \\ 
R2Gen \citep{chen2020generating} & 2020 & 0.165 & 0.371 & 0.187 \\
PPKED \citep{liu2021exploring} & 2021 & 0.168 & 0.376 & 0.190 \\
VTI \citep{najdenkoska2021variational} & 2021 & 0.154 & 0.375 & \underline{0.218} \\
SGF \citep{li2022self} & 2022 & \textbf{0.215} & \underline{0.415} & 0.201 \\ 
CMCL \citep{liu2022competence} & 2022 & 0.162 & 0.378 & 0.186 \\
R2GenCMN \citep{chen2022cross} & 2022 & 0.168 & 0.370 & 0.198 \\
CMM+RL \citep{qin2022reinforced} & 2022 & 0.181 & 0.384 & 0.201 \\
JPG \citep{you2022jpg} & 2022 & 0.174 & 0.377 & 0.193 \\ 
PGT \citep{yan2022prior} & 2022 & 0.181 & 0.381 & 0.203 \\ 
ITA \citep{wang2022inclusive} & 2022 & 0.188 & 0.382 & 0.208 \\ 
CvT-DistilGPT2 \citep{nicolson2023improving} & 2023 & 0.177 & 0.377 & 0.193 \\
DCL \citep{li2023dynamic} & 2023 & 0.163 & 0.383 & 0.193 \\
UAR \citep{li2023unify} & 2023 & 0.200 & 0.405 & \underline{0.218} \\
R2GenGPT~\citep{wang2023r2gengpt} & 2023 & 0.173 & 0.377 & 0.211\\
METransformer~\citep{wang2023metransformer} & 2023 & 0.172 & 0.380 & 0.192\\
Med-LLM~\citep{liu2024context} & 2024 & 0.168 & 0.381 & 0.209\\
MAN~\citep{shen2024automatic} & 2024 & 0.170 & 0.386 & 0.213\\
MCSAM \citep{tao2024memory} & 2024 & 0.184 & 0.394 & 0.210 \\
ASGMD~\citep{xue2024generating} & 2024 & 0.173 & 0.397 & 0.206\\
MAMRG~\citep{shen2024automatic} & 2024 & 0.170 & 0.386 & 0.213\\
MiniGPT-4-RRG~\citep{liu2024bootstrapping} & 2024 & 0.184 & 0.390 & 0.208\\
MDAKF~\citep{tan2024medical} & 2024 & 0.174 & 0.389 & 0.194 \\
SSM~\citep{zhang2025automatic} & 2025 & 0.201 & 0.411 & 0.211\\
VLCI~\citep{chen2025cross} & 2025 & 0.189 & 0.397 & 0.204\\
CAT~\citep{tang2025cross} & 2025 & 0.176 & 0.383 & 0.195\\
LACCOL~\citep{liu2025label} & 2025 & 0.172 & 0.392 & 0.198\\
R2gen-mamba~\citep{sun2025r2gen} & 2025 & 0.176 & 0.382 & 0.208\\
MWCL~\citep{usman2025mwcl} & 2025 & 0.196 & 0.409 & 0.213\\
\hline
LLaMA-XR (this study) & - & 0.180 & \textbf{0.433} & \textbf{0.336} \\ 
$\Delta$ SOTA & - & & (+4.34\%) & (+54.13\%)\\
\hline 

\hline
\end{tabular} 
\end{table}

Previous research has demonstrated that METEOR is more effective than BLEU in capturing semantic accuracy and linguistic fluency, particularly in specialized domains where precise terminology and structured language are critical, such as medical reporting~\citep{banerjee2005meteor}. In a similar vein, ROUGE-L has been widely recognized as a more suitable metric for assessing textual coherence and semantic fidelity, especially in tasks involving summarization and natural language generation within healthcare contexts~\citep{lin2004rouge}. Given that medical reports demand factual correctness, structured completeness, and linguistic coherence, the use of METEOR and ROUGE-L provides a more clinically meaningful and domain-appropriate evaluation of model performance than surface-level n-gram matching metrics.

To rigorously assess the performance of our model and highlight that, despite a marginally lower BLEU-4 score, our ROUGE-L and METEOR scores reflect superior linguistic coherence and clinical relevance, we conducted a comparative analysis against CvT-DistilGPT2~\citep{nicolson2023improving}, a model incorporating a pretrained GPT architecture with a substantial parameter count. The results of this qualitative performance comparison between our proposed LLaMA-XR framework and CvT-DistilGPT2 are presented in Table~\ref{tab:my_label1}.
Although LLaMA-XR yields a slightly lower BLEU-4 score, it achieves notable improvements in both ROUGE-L and METEOR. This distinction is particularly important in the domain of medical report generation, where lexical overlap alone, as measured by BLEU-4, is often inadequate for evaluating the semantic integrity and clinical accuracy of generated text~\citep{nguyen2023pragmatic}. These findings underscore the relevance of using metrics that prioritize meaningful content alignment over surface-level word matching when assessing models in the clinical NLP domain.

\begin{table}[!htp]
    \centering
    \caption{ Performance analysis of the LLaMA-XR-generated report with CvT-DistilGPT2. 
    }
    \label{tab:my_label1}
    \small
    \begin{tabular}{p{1.65cm}|p{3cm}|p{3cm}|p{3cm}|p{4cm}}
    \hline
     \multicolumn{5}{c}{Input X-ray image with attention map} \\
     \multicolumn{5}{c}{\includegraphics[width=.2\textwidth]{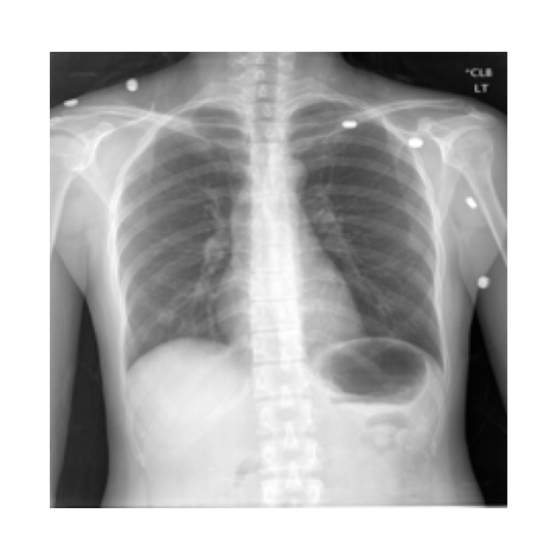}  \includegraphics[width=.2\textwidth]{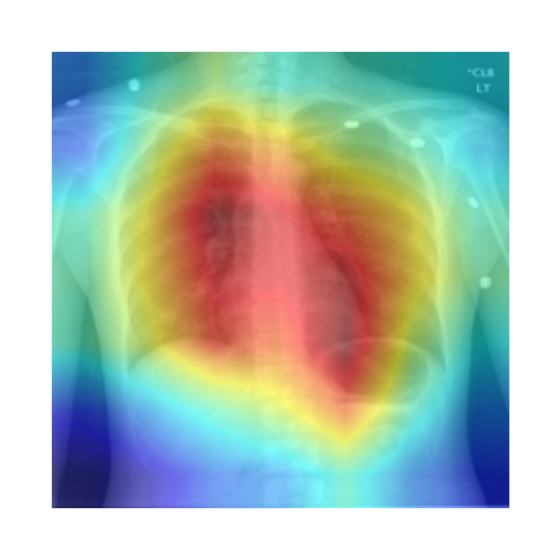}   \includegraphics[width=.2\textwidth]{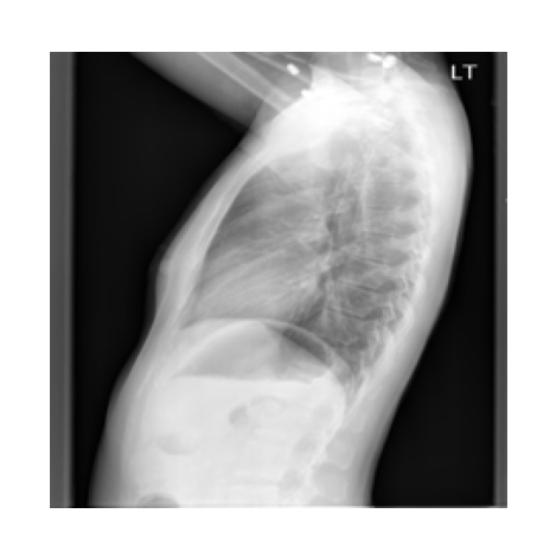}  \includegraphics[width=.2\textwidth]{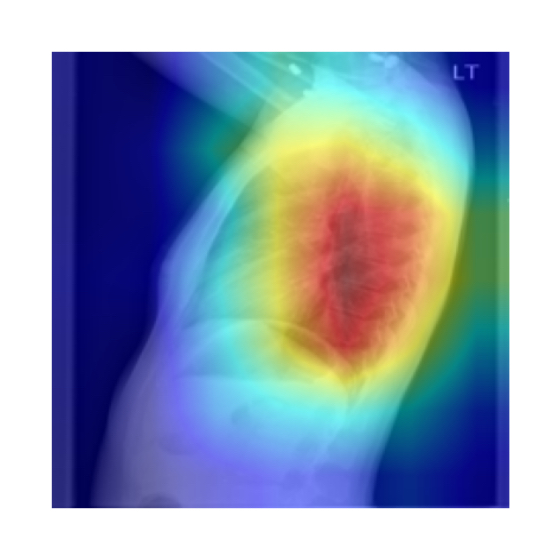}} \\
     \hline
     Reference report & \multicolumn{4}{p{14cm}}{The heart and mediastinum are unremarkable. The lungs are clear without infiltrate. There is no effusion or pneumothorax.}\\\hline
     CvT-DistilGPT2 & \multicolumn{4}{p{14cm}}{The lungs are clear bilaterally. Specifically no evidence of focal consolidation pneumothorax or pleural effusion. Cardiomediastinal silhouette is unremarkable. Visualized osseous structures of the thorax are without acute abnormality.}\\ \hline
     LLaMA-XR & \multicolumn{4}{p{14cm}}{The heart size and mediastinal contours appear within normal limits. The lungs are clear. There is no focal airspace opacity. No pleural effusion or pneumothorax is seen.}\\
\hline\hline
     \makecell[l]{Category} & \makecell[c]{Reference report}  &CvT-DistilGPT2~\citep{nicolson2023improving} & \makecell[c]{LLaMA-XR}& \makecell[c]{Remark}\\
     \hline
     Performance & \makecell[c]{-} & \begin{tabular}{@{}ccc@{}} BL4 & RL & MRT \end{tabular} & \begin{tabular}{@{}ccc@{}} BL4 & RL & MRT \end{tabular}  & \makecell[c]{-}\\
     & & \begin{tabular}{@{}ccc@{}} 0.1 & 0.25 & 0.29\end{tabular} & \begin{tabular}{@{}ccc@{}} 0.07 & 0.57 & 0.5\end{tabular} & \\
    \hline
     Heart & The heart and mediastinum are unremarkable.  & The cardiomediastinal silhouette is unremarkable. & The heart size and mediastinal contours appear within normal limits. & LLaMA-XR provides more precise anatomical details.\\
     \hline
     Lungs & The lungs are clear without infiltrate.  & The lungs are clear bilaterally. & The lungs are clear. & LLaMA-XR is more concise, whereas CvT-DistilGPT2 introduces redundancy.\\
     \hline
     Focal Airspace Opacity & Not explicitly mentioned.  & No focal consolidation.& There is no focal airspace opacity. & Both models identified a clinically significant feature that was not explicitly mentioned in the reference report.\\
     \hline
     Pleural Effusion & There is no effusion.  & No pleural effusion. & No pleural effusion is seen. & Both models are similar, but LLaMA-XR maintains structured phrasing.\\
     \hline
     Pneumothorax & There is no pneumothorax.  & No pneumothorax. & No pneumothorax is seen. & Both models are similar, but LLaMA-XR is more structured.\\
     \hline
    \end{tabular}
\end{table}



Our LLaMA-XR framework consistently outperforms CvT-DistilGPT2 on both METEOR and ROUGE-L, underscoring its superior capability in generating semantically accurate and coherent medical texts. The METEOR metric, which accounts for synonymy, stemming, and semantic matching, is particularly well-suited for the medical domain, where multiple terminologies may describe the same condition or finding. In contrast, ROUGE-L evaluates the longest common subsequence (LCS), making it highly effective for assessing the preservation of structured medical content, such as causal relationships and anatomical consistency~\citep{lin2004rouge}.
Although CvT-DistilGPT2 achieves a higher BLEU-4 score, its outputs tend to prioritize lexical overlap over true semantic alignment, often producing more generic reports that lack the specificity and clinical precision required in radiological interpretation. These results highlight the limitations of relying solely on BLEU-based metrics and emphasize the importance of clinically informed evaluation frameworks in medical text generation.
\section{Discussion}

\subsection{Qualitative Analysis}
Table~\ref{fig:comparison} presents a qualitative comparison between reports generated by CvT-DistilGPT2 and our proposed LLaMA-XR framework. The results reveal that LLaMA-XR consistently produces reports with enhanced linguistic coherence and clinical relevance. In contrast, CvT-DistilGPT2~\citep{nicolson2023improving} frequently introduces extraneous content not present in the reference report (highlighted in red), which may compromise the reliability of its outputs in clinical settings due to the risk of hallucinated or inaccurate findings. However, it is worth noting that when such additional observations are clinically accurate, they could potentially offer supplementary value to clinicians by highlighting findings that may have been overlooked in the original interpretation. Nevertheless, distinguishing between helpful augmentation and unreliable hallucination remains a critical challenge for safe and trustworthy AI deployment in medical contexts.

LLaMA-XR demonstrates a higher degree of faithfulness to the source content, effectively retaining key clinical details such as normal heart size and pulmonary vascularity. Additionally, it includes accurate findings (highlighted in green) that are overlooked by CvT-DistilGPT2, underscoring its improved capacity to capture and preserve essential medical information. This distinction is particularly evident in the second example, where LLaMA-XR correctly identifies the absence of pneumonia, pleural effusion, and pulmonary edema, which are critical findings for accurate clinical interpretation.
These observations highlight the robustness and clinical precision of LLaMA-XR, demonstrating its potential to generate reliable, non-hallucinated, and clinically actionable radiology reports. The ability to minimize hallucinations while preserving nuanced medical information is crucial for the safe deployment of AI systems in healthcare environments, where diagnostic accuracy and patient safety are paramount.

\begin{table}[!ht]
\centering
\caption{Comparison of reference and generated reports from CvT-DistilGPT2 and LLaMA-XR. Findings correctly generated by both models are shown in normal black text (not highlighted). Additional findings not present in the original report are highlighted in red. Correct findings missed by one model but captured by the other are highlighted in green.}
\label{fig:comparison}
\begin{tabular}{lp{4cm}p{4.5cm}p{4.5cm}}
\hline
Image & Reference & CvT-DistilGPT2~\citep{nicolson2023improving} & LLaMA-XR (this study)\\ \hline\hline

\raisebox{-.95\height}{\includegraphics[width=0.1\textwidth]{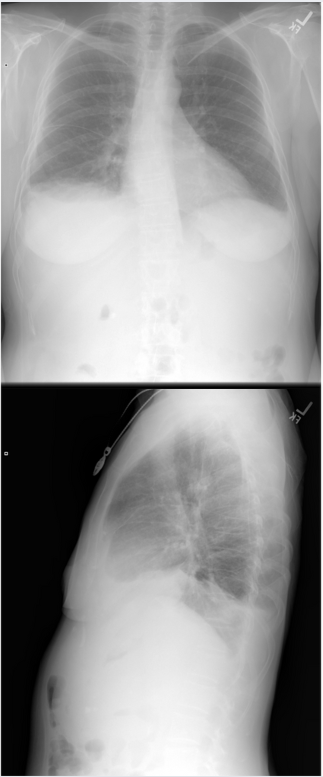}} & 
The lungs and pleural spaces show no acute abnormality. Heart size and pulmonary vascularity within normal limits. & 
The cardiomediastinal silhouette and pulmonary vasculature are within normal limits in size. The lungs are clear of focal airspace disease pneumothorax or pleural effusion. \sethlcolor{red!40}\hl{There are no acute bony findings.}& 
The lungs are clear. The heart is normal in size and contour. The pulmonary vascularity is within normal limits in appearance.\\ \hline

\raisebox{-1\height}{\includegraphics[width=0.1\textwidth]{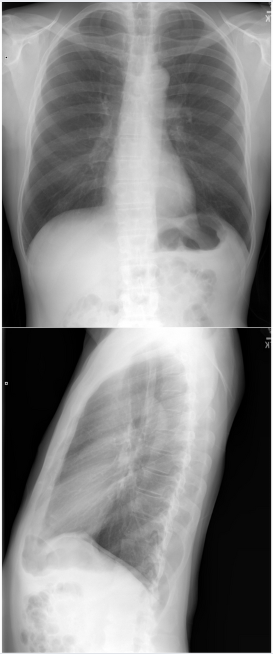}} & 
Heart size normal. Lungs are clear. XXXX are normal. No pneumonia, effusions, edema, pneumothorax, adenopathy, nodules or masses. & 
The lungs are clear bilaterally. Specifically no evidence of \sethlcolor{red!40}\hl{focal consolidation} pneumothorax or pleural effusion. Cardiomediastinal silhouette is unremarkable. Visualized osseous structures of the thorax are without acute abnormality. & 
Heart size normal. Lungs are clear. There is no evidence of \sethlcolor{green!40}\hl{pneumonia}, pleural effusion, pulmonary edema, pneumothorax, \sethlcolor{red!40}\hl{lymphadenopathy}, nodules, or masses.\\ \hline

\hline
\end{tabular}
\end{table}

Table~\ref{fig:comparison-2} further reinforces the superiority of LLaMA-XR by presenting a qualitative comparison with the PGT model. The analysis reveals that LLaMA-XR consistently generates reports with greater fluency, completeness, and clinical accuracy. While PGT occasionally produces correct descriptions, it often omits critical medical findings that are present in the reference reports, thereby limiting its clinical reliability.
For example, in the first case, PGT fails to mention the presence of focal consolidation, a key diagnostic feature that LLaMA-XR correctly identifies. In the second case, LLaMA-XR goes further by incorporating additional relevant observations, such as bronchovascular crowding, thereby enhancing the diagnostic completeness of the report.
These findings highlight LLaMA-XR’s improved ability to generate comprehensive and clinically aligned narratives, which is especially vital in medical reporting where factual precision and diagnostic thoroughness are paramount. This analysis reaffirms the robustness of LLaMA-XR in delivering expert-level interpretations, positioning it as a reliable tool for supporting clinical decision-making.

\begin{table}[!ht]
\centering
\caption{Comparison of reference and generated reports from PGT and LLaMA-XR. Findings correctly generated by both models are shown in normal black text (not highlighted). Additional findings not present in the reference report are highlighted in red. Correct findings missed by one model but identified by the other model are highlighted in green.}
\label{fig:comparison-2}
\begin{tabular}{lp{4cm}p{4.5cm}p{4.5cm}}
\hline
Image & Reference & PGT~\citep{yan2022prior} & LLaMA-XR (this study)\\ \hline\hline

\raisebox{-.8\height}{\includegraphics[width=0.1\textwidth]{ca.PNG}} & 
Lungs are clear without focal consolidation, effusion, or pneumothorax. Normal heart size. Bony thorax and soft tissue is unremarkable.& 
The lungs are clear. There is no pleural effusion or pneumothorax. \sethlcolor{green!40}\hl{The heart} and  \sethlcolor{red!40}\hl{mediastinum} \sethlcolor{green!40}\hl{are normal}. The bony thorax are normal.& 
The lungs are clear. There is \sethlcolor{green!40}\hl{no focal consolidation}, effusion, or pneumothorax. Bony thorax unremarkable. \\ \hline

\raisebox{-1.1\height}{\includegraphics[width=0.1\textwidth]{cb.PNG}} & 
Low lung volumes bilaterally, with lungs otherwise grossly clear. No focal consolidation, pneumothorax, or large pleural effusion. The cardiomediastinal silhouette is unremarkable. No acute osseous abnormalities identified. & 
Lung volumes are low. Specifically no evidence of focal consolidation pneumothorax or pleural effusion. Cardiomediastinal silhouette is unremarkable. Visualized osseous structures of the thorax are without acute abnormality.& 
Low lung volumes bilaterally with \sethlcolor{red!40}\hl{central bronchovascular crowding} without focal consolidation, pleural effusion, or pneumothorax. Cardiomediastinal silhouette is unremarkable. No acute osseous abnormality.\\ \hline

\hline

\end{tabular}
\end{table}

\subsection{Findings and Observations}
In this study, several key findings and observations were made regarding the use of various models for medical report generation. The fine-tuning of LLaMA 3.1~\citep{dubey2024llama}, a large autoregressive decoder-only model with 8 billion parameters, has demonstrated strong performance in generating medical reports. This model benefits from being trained on a substantial amount of diverse data, allowing it to effectively capture the patterns required for generating clinically relevant text. Another important observation is that classification models such as DenseNet-121~\citep{cohen2022torchxrayvision}, when used as an embedding, can achieve excellent performance in generating medical reports when provided with clear and meaningful instructions. In this study, we used a pre-trained version of DenseNet-121, fine-tuned on the same domain of medical imaging data, which likely contributed to its strong performance. This pre-training allows the model to better understand the specific nuances of medical images, enhancing its ability to interpret and generate relevant medical text. In addition, the use of Unsloth, which supports QLoRA~\citep{dettmers2024qlora} 4-bit quantization, has significantly improved the efficiency of the report generation process in terms of both speed and memory consumption. Unsloth's ability to perform quantization reduces the model's memory requirements, enabling faster processing without compromising the quality of the generated reports.

\subsection{Limitations and Future Work}
While our study utilized IU-Xray~\citep{demner2016preparing}, one of the two widely recognized benchmark datasets, we selected it due to its high relevance and suitability for the research objectives. Its well-curated structure and established use in prior work make it a robust foundation for evaluating our approach. Nonetheless, future work could benefit from extending the analysis to the MIMIC-CXR~\citep{johnson2019mimic} dataset to further validate and generalize the findings.

Among the various pre-trained models available for image embedding, we employed DenseNet-121 due to its demonstrated strong performance and efficient parameter utilization in medical imaging tasks. The choice reflects a balance between computational efficiency and representational power. Importantly, our framework is designed to be modular and architecture-agnostic, allowing for seamless integration of alternative backbone models such as ResNet, EfficientNet, or Vision Transformers. Future studies may explore these and other architectures within our framework to assess their comparative effectiveness and adapt the model to diverse application requirements.

Additionally, we focused exclusively on QLoRA as our fine-tuning method, chosen for its efficiency and effectiveness in resource-constrained settings. Future work may investigate other fine-tuning strategies, including LoRA, Prefix Tuning, or Full Parameter Fine-Tuning, to assess their relative impact and broaden the applicability of our approach across various model sizes and domains.

From a clinical radiology perspective, the integration of LLaMA-XR into radiologist workflows requires further exploration, particularly in terms of real-world deployment and validation~\citep{kim2024large}. While the framework demonstrates improved semantic and linguistic performance, its clinical utility hinges on how it supports or augments human interpretation~\citep{dikici2020integrating}. For instance, LLaMA-XR could be deployed to draft preliminary reports for radiologist review in settings with limited specialist availability, such as rural or resource-constrained environments~\citep{guo2024can}. The model's attention heatmaps may support transparency and clinician trust by visually linking predictions to image regions of interest~\citep{watanabe2022improving}, though their practical interpretability in diagnostic workflows still needs to be assessed. Since LLaMA-XR generates narrative reports, future development should explore alignment with structured reporting frameworks such as RSNA templates or BI-RADS to enhance consistency, completeness, and data interoperability~\citep{granata2022structured}. Additionally, broader validation using clinical datasets, including cases with incidental findings, medical devices, or less common pathologies, is essential for robust performance~\citep{ahluwalia2023subgroup}. Ongoing radiologist feedback and real-world evaluation will be critical to ensure safety, relevance, and integration into clinical practice~\citep{dikici2020integrating}.

Building on these system-level considerations, further validation should also focus on the clinical content and safety of the generated reports. While ROUGE-L and METEOR provide insight into semantic and syntactic quality, they do not evaluate whether key clinical findings are correctly identified or whether the reports omit critical abnormalities or introduce hallucinated content~\citep{adams2023leveraging, ahluwalia2023subgroup}. A small-scale qualitative review by radiologists, such as case-based annotations comparing generated reports to reference standards, could offer valuable insight into clinical accuracy, safety, and interpretability~\citep{guo2024can}. This type of human evaluation would help determine whether the model supports diagnostic decision-making or introduces risks through omission or irrelevant information~\citep{dikici2020integrating}.

In addition, the model's ability to recognise uncommon or incidental findings requires further exploration. Chest radiographs frequently contain incidental nodules, bony lesions, or devices such as central lines and pacemakers, which may be underrepresented in training datasets. These elements are clinically important but can be overlooked by automated systems that rely on frequent or highly visible patterns~\citep{ahluwalia2023subgroup}. Understanding how LLaMA-XR handles such low-prevalence features is critical for deployment in real-world practice. Future studies should include cases with atypical pathologies and radiological devices to assess whether the model can accurately identify, interpret, or appropriately omit these findings based on clinical context~\citep{dikici2020integrating,granata2022structured}.

\section{Conclusions}
In this study, we proposed LLaMA-XR, a novel framework for automated radiology report generation that combines DenseNet-121 for visual feature extraction with LLaMA 3.1 for text generation and structured text completion. By leveraging multi-view X-ray embeddings and prompt-based conditioning, our model bridges the semantic gap between visual and textual modalities more effectively than previous approaches.
LLaMA-XR achieved state-of-the-art performance on the IU X-ray dataset, attaining a ROUGE-L score of 0.433 and a METEOR score of 0.336, indicating its ability to generate semantically accurate and clinically coherent reports. Although BLEU scores were marginally lower than those of some recent models, they remained competitive, suggesting strong linguistic fluency, semantic alignment, and syntactic structure in the generated outputs.
The use of structured prompts and optimized fine-tuning strategies contributed to enhanced report quality and computational efficiency. This underscores the effectiveness of our framework in producing reliable medical narratives with reduced memory and processing overhead.
Overall, our findings demonstrate the potential of combining state-of-the-art vision and language models to advance AI-assisted radiology, contributing to more accurate, efficient, and clinically useful diagnostic tools.

\section*{CRediT authorship contribution statement}
\textbf{Md. Zihad Bin Jahangir}: Conceptualization, Methodology, Investigation, Software, Visualization, Writing – original draft. 
\textbf{Muhammad Ashad Kabir}: Conceptualization, Methodology, Resources, Writing – original draft, Writing – review \& editing, Validation, Visualization, Supervision.
\textbf{Sumaiya Akter}: Conceptualization, Visualization, Writing – original draft. \textbf{Israt Jahan}: Conceptualization, Visualization, Writing – original draft.
\textbf{Minh Chau}: Validation, Writing – review \& editing.

\section*{Data availability statement}
The study used a secondary dataset available at \url{https://openi.nlm.nih.gov/faq?download=true}. The source code is available at \url{https://github.com/zihadbinjahangir/LLaMA-XR}. 

\section*{Declaration of competing interest}
The authors declare that they have no conflict of interest.

\section*{Funding}
This research did not receive any specific grant from funding agencies in the public, commercial, or not-for-profit sectors.

\section*{Ethics statement}
Not applicable.

\section*{Consent for publication}
Not applicable.

\section*{Acknowledgments}
None.

\bibliographystyle{elsarticle-num-names}
\bibliography{reference}

\end{document}